\documentclass[twocolumn,aps,pra,longbibliography]{revtex4-2}
\usepackage{amsmath}
\usepackage{amssymb}
\usepackage{xfrac}
\usepackage{graphicx}
\usepackage[utf8]{inputenc}
\usepackage[T1]{fontenc}
\usepackage{color}
\newcommand{\av }[1]{\braket{#1}}
\renewcommand{\i}{\mathrm{i}}
\usepackage{braket}

\begin{document}

\title{Bound states in the continuum in a fluxonium qutrit}

\author{María Hita-Pérez$^\dagger$, Pedro Orellana$^\ddag$, Juan José García-Ripoll$^\dagger$, Manuel Pino$^{*\dag}$}
\address{$\dagger$ Institute of Fundamental Physics IFF-CSIC, Calle Serrano 113b, Madrid 28006, Spain\\
$\ddag$ Departamento de Física Universidad Técnica Federico Santa Maríaa,Casilla 110-V, Valparaíso, Chile.\\
$^*$ Nanotechnology Group, USAL-Nanolab,
Universidad de Salamanca, E-37008 Salamanca, Spain.
}

\begin{abstract} 
The heavy fluxonium at zero external flux has a long-lived state when coupled capacitively to any other system. We analyze it by projecting all the fluxonium relevant operators into the qutrit subspace, as this long-lived configuration corresponds to the second excited fluxonium level. This state becomes a bound-state in the continuum (BIC) when the coupling occurs to an extended system supporting a continuum of modes.  In the case without noise, we find BIC lifetimes that can be much larger than seconds $T_1\gg {\rm s}$ when the fluxonium is coupled to a superconducting waveguide, while typical device frequencies are in the order of ${\rm GHz}$. We have performed a detailed study of the different sources of decoherence in a realistic experiment, obtaining that upward transitions caused by a finite temperature in the waveguide and decay induced by $1/f$-flux noise are the most dangerous ones. Even in their presence, BICs decay times could reach the range of ${T_1\sim \rm 10^{-1} ms},$ while preparation times are of the order of $10^{2}$ns. 
\end{abstract}
\maketitle

\section{Introduction}

Confined quantum excitations generally decay when coupled to a band of states with a continuous spectrum~\cite{dirac1927}. There are some exceptions to those decay processes where a confined state lying at the continuum part of the spectrum lives forever. Those bound states in the continuum (BICs) were predicted long ago by von Neumann and Wigner~\cite{von1993}. The BICs have appeared on several platforms, some following the laws of quantum mechanics as solid-state devices~\cite{capasso1992observation,ramos2019,de2006,gonzalez2010bound}, some others---under the wave-particle duality---obeying classical wave mechanics~\cite{parker1966resonance, parker1967resonance}. For instances, there have been many studies of BIC's in photonic devices \cite{weimann2013,corrielli2013,yang2014,ovcharenko2020,kodigala2017,yu2020acousto} since their first experimental observation around ten years ago~\cite{plotnik2011}.

Besides the importance of BICs from a fundamental point of view, those states have also found a broad range of applications in lasing, light trapping, and sensing, among others\ \cite{chen2022}. For instance,  Huang \emph{et al.} have employed the concept of a supercavity mode created by merging symmetry-protected and accidental BICs in the momentum space and realizing an efficient laser based on a finite-size cavity\ \cite{hwang2021ultralow}. BICs have applications as high-Q building blocks for acoustic sensors, antennas, and topological acoustic structures\ \cite{deriy2022bound}. Another example of the technological utility of BICs is the work of Mao \emph{et al.}, where quasi-BIC magnetic resonance has been shown to improve the chiral lateral force on the paired enantiomers with linearly polarized illumination\ \cite{mao2022sieving}.

There are also proposals to study BICs in high-coherence quantum optical devices. Inspired by the physics of classical BICs with confined electromagnetic fields~\cite{Dong2009,tufarelli2013dynamics,tufarelli2014non,hoi2015probing,cotrufo2019excitation,Zhou2008}, recent works proposed to use two-level systems or qubits~~\cite{ calajo2019exciting,feiguin2020qubit} to create an extremely long-lived and confined single-photon excitation. Another approach based on two-level emitters is to employ their collective photon-mediated interactions to create extended BIC states---authors refer to as a multi-dark state---that live on two or more separated qubits~\cite{zanner2022}.

\begin{figure}[t!]
\includegraphics[width=0.9\columnwidth]{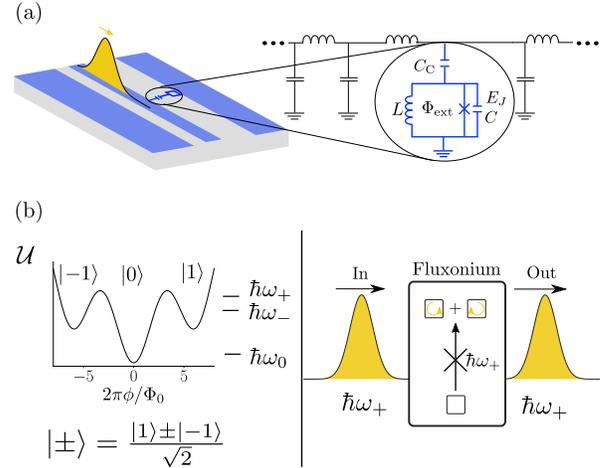}
\caption{  (a) A fluxonium threaded by an external flux $\Phi_\text{ext}$ is capacitively coupled to a superconducting waveguide. (b) The device potential at $\Phi_\text{ext}=0$ as a function of fluxonium phase $2\pi\phi/\Phi_0,$ with $\Phi_0$ the magnetic flux quantum. Its three minima states are denoted by $\ket{-1},\ket{0},\ket{1}.$  The first and second excited eigenstate of the fluxonium are antisymmetric and symmetric superposition of those states denoted by $\ket{-},\ket{+},$ having each one odd and even parity under fluxonium flux reversal. (c) Due to the odd parity of the capacitive coupling, incoming radiation at the frequency of the $\ket{+}$ state cannot excite this state. This state becomes completely uncoupled from the waveguide if the device capacitance is made large enough.}
\label{Fig:setup}
\end{figure}

In this work, we show how to engineer a scalable, compact BIC using a superconducting circuit, a fluxonium qutrit~\cite{manucharyan2009fluxonium}, capacitively \textit{embedded in the continuum of microwave excitations} from a co-planar waveguide [cf. Fig.~ \ref{Fig:setup}(a)]. Similar to the classical setup in Ref.~\cite{ahumada2018bound}, where the BIC lives in a photonic resonator connected to an open waveguide, our BIC is the confined plasmonic excitation that lives in the fluxonium loop and is prevented from decaying into the microwave guide. More precisely, the BIC state is the symmetric excited state $\ket{+}$ of the fluxonium potential at zero external flux $\Phi_\text{ext}=0$ [cf. Fig~\ref{Fig:setup}(b)]. This state is a BIC state because the $\ket{+}\to\ket{0}$ transitions are suppressed because the charge operator is antisymmetric and cannot connect both states, as experimentally observed in Ref.~\cite{manucharyan2009fluxonium}. Moreover, the $\ket{+}\to\ket{-}$ transition can also be suppressed by a suitable choice of the capacitive to inductive energy ratios $E_\text{J}/E_\text{C}$, as in capacitively shunted flux qubits~\cite{hita2021ultrastrong, hita2021three} or capacitively shunted heavy fluxonium qubits at half frustration~\cite{earnest2018,Zhang2021}. Using these design considerations and realistic parameters, we prove that the BIC can reach very long lifetimes, larger than seconds in the noise-less system and $10^-1$ms in realistic experiments. 

The structure of this work is as follows. We begin with a thorough study of the fluxonium qutrit and the effective capacitive and inductive interactions between the qutrit and external fields. This study reveals a symmetry point at which the $\ket{+}$ state effectively decouples from all external fields. When the fluxonium is placed in a microwave guide, the capacitive interaction between the qutrit and the propagating fields is described by the effective models we found, and the $\ket{+}$ state becomes a BIC. The lifetime of this state is shown to be highly long even in the presence of realistic $1/f$-flux noise, a finite temperature in the waveguide, dielectric and inductive losses. Finally, we also discuss some applications of these compact BICs in quantum information and quantum sensing, including open questions such as the robust preparation of the $\ket{+}$ in the open system. Those applications suggest that BICs may have a technological impact in quantum electrodynamics with superconducting circuits similar to the one they had in photonic science.

\section{Fluxonium Qutrit}

Let us formalize the intuitive picture of a BIC using a fluxonium circuit. However, first, we must show that a field coupled capacitively to the fluxonium cannot excite transitions in and out of the second excited. This information is obtained from the expansion of the charge operator and the Hamiltonian in the relevant low-energy subspace. This subspace has a qutrit structure, where the BIC mode corresponds to the second excited state, $\ket{+}$. Incidentally, in this qutrit representation of the heavy fluxonium states, flux and charge operators adopt the simple representation of $S_x$ and $S_y$ spin-1 operators, respectively.

\begin{figure}[t!]
\includegraphics[width=0.9\columnwidth]{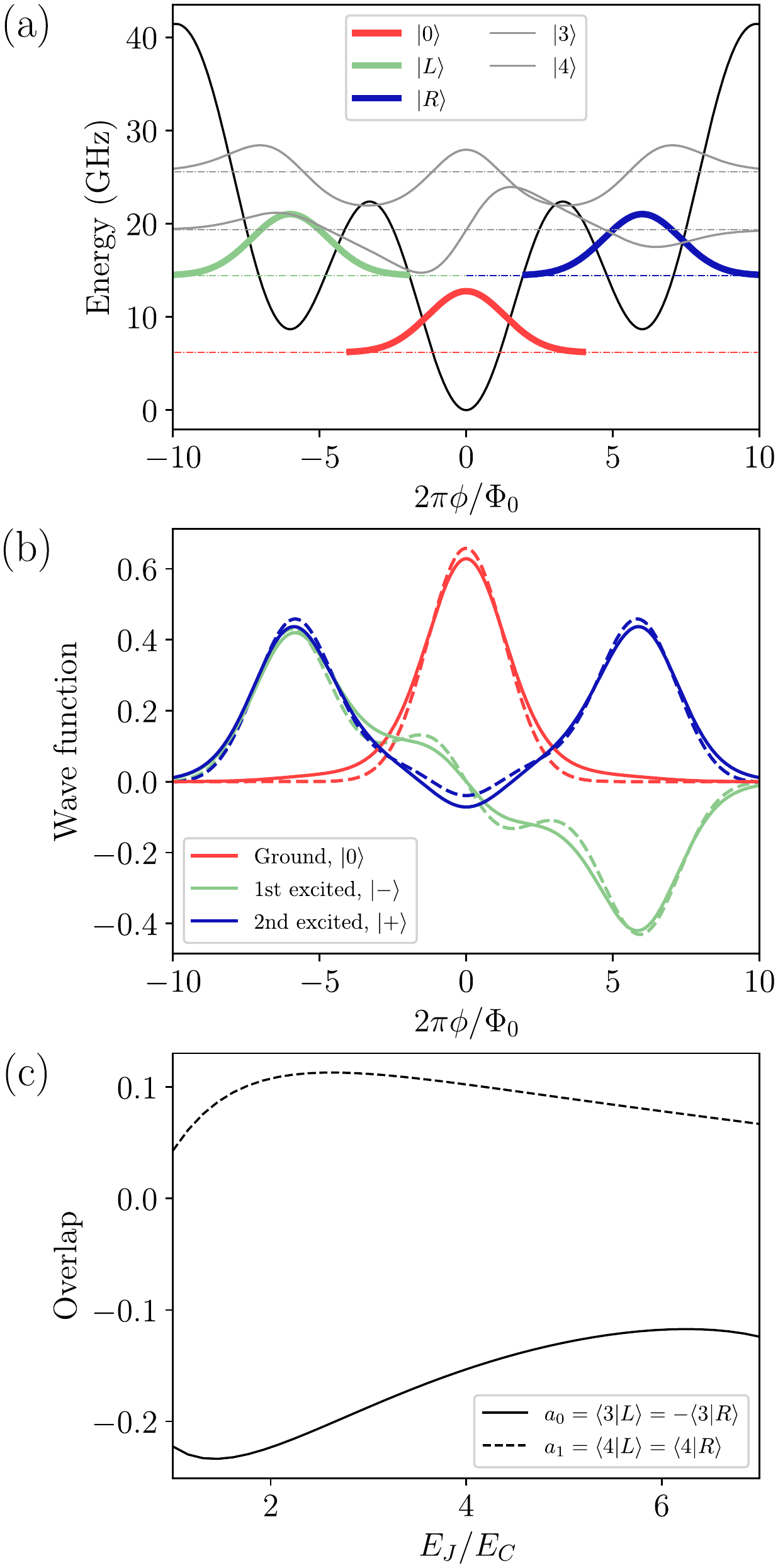}
\caption{(a) Potential energy of the fluxonium for $E_J = 10$ GHz,
 $E_C = 3.6$ GHz and $E_L = 0.46$ GHz~\cite{vool2014non} as solid black line as a function of fluxonium phase $2\pi \phi/\Phi_0$. Gaussian states in the left $\ket{L}$, center $\ket{0}$ and right $\ket{R}$ wells  are denoted with thick solid lines, while thin gray ones are used for  third and fourth excited eigenstates. The dashed lines indicate the energy of each of the wavefunctions (Gaussian potential approximation for the Gaussian states). (b) Wavefunction amplitudes of the three lowest eigenstates (solid line) and the approximated ones considering the Gram-Schmidt orthogonalization up to the $4^{th}$ excited state (dashed line) for the previous fluxonium characteristic energies. (c) Overlap between the Gaussian states in the left $\ket{L}$ and right $\ket{R}$ wells of the potential and the excited states considered for the Gram-Schmidt orthogonalization, $3^{th}$ (solid line) and $4^{th}$ (dashed line)excited states, as a function of $E_J/E_C$.}
\label{Fig:eigenstates}
\end{figure}

A fluxonium~\cite{manucharyan2009fluxonium} consists of a single Josephson junction with Josephson energy $E_J$ shunted by a capacitance $C_f$ and a large inductance $L$, as shown in Fig.~\ref{Fig:setup}(a). The Hamiltonian for such a system is given by:
\begin{align}
    H_f(q,\phi)=&\frac{1}{2C_f}q^2+  \mathcal{U}(\phi),\mbox{ with} \nonumber \\
    \mathcal{U}(\phi)=&\frac{1}{2L}\phi^2-E_J\cos{\left(2\pi \frac{\phi+\Phi_\text{ext}}{\Phi_0}\right)}.
\end{align}
Here, $q$ is the charge difference on the capacitance, $\phi$ is the conjugate flux operator, and $\Phi_\text{ext}$ is the external flux passing through the superconducting loop. We work at $\phi_{\rm ext}=0$ so that the potential has the shape depicted in 1(b). This is not the usual working point of the fluxonium qubit\ \cite{nguyen2019}, which is  usually operated at $\Phi_\text{ext}=\Phi_0/2$ so that it presents a double well potential.

The characteristic energies of the fluxonium are the junction's Josephson energy $E_J$, the charging energy introduced by the capacitance $E_C=e^2/2C_f$, and the inductive energy introduced by the inductance $E_L=(\hbar/2e)^2/L$. The main difference between this and other inductively shunted Josephson junction devices lies precisely in the relation between these parameters, which satisfy $E_L\ll E_J$ and $1\lesssim E_J/E_C$~\cite{peruzzo2021geometric}. The heavy fluxonium is realized approximately for $E_J/E_C>5$~\cite{Zhang2021}. 

Let us describe an analytical derivation of the qutrit Hamiltonian and relevant operators. At the symmetry point $\Phi_\text{ext} =0$, the potential energy of the fluxonium has two local minima on both sides of the global one. The lowest energy eigenstates around the three minima are denoted $\ket{L},\ \ket{0},\ \ket{R},$ as depicted in Fig.~\ref{Fig:eigenstates}(a). Intuitively, one would want to use $\ket{L},\ \ket{0},\ \ket{R}$ as a qutrit basis. However, the use of the $\ket{L}$ and $\ket{R}$ vectors is problematic in common situations where they have a strong overlap and are close in energy to nearby excitations (upper thin gray lines in panel (a) of Fig.~\ref{Fig:eigenstates}).

One solution is to replace the $\ket{L}$ and $\ket{R}$ states with slightly modified vectors that have been orthogonalized with respect to other low-energy excitations. Fig.~\ref{Fig:eigenstates}(b) compares the exact eigenstates (solid lines) of a fluxonium with intermediate values of $E_J/E_C$ together with the approximated eigenstates (dashed lines) computed with a Gram-Schmidt orthogonalization up to the $4^{th}$ excited state.  The coefficients involved in this orthogonalization are presented in Fig.\ \ref{Fig:eigenstates}(c) as a function of $E_J/E_C.$ Notice that the maximum in the coefficient $a_0$ is due to the avoided level crossing between the second and third excited state, so for larger values of $E_J/E_C$ our picture based on an isolated state in each potential well is no longer valid.  In summary, the agreement between exact and approximated eigenstates is good. It helps capture the part of the excited wavefunction that tunnels to the intermediate region $\phi\simeq 0$, a feature that is not present in the original intuitive expansion. Moreover, these features introduce relevant qutrit interaction terms that are mediated by higher energy excitations.

\begin{figure}[t!]
\includegraphics[width=1\columnwidth]{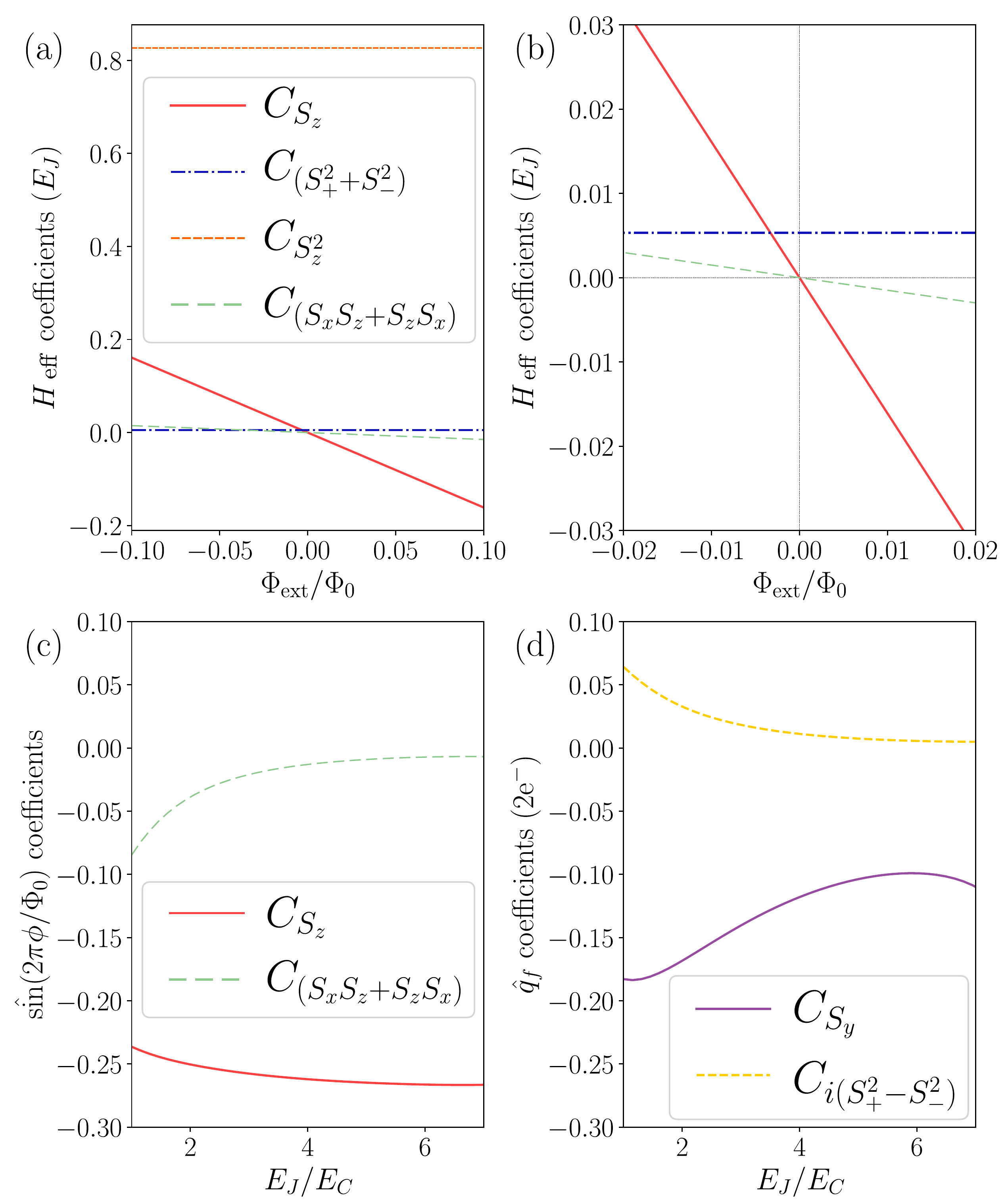}
\caption{{Expansion of effective Hamiltonian $H_{\rm eff},$ sine of flux $\sin(2\pi\phi/\Phi_0)$ and
charge $q$ in a generic form $L=\sum_{i} C_{o_i}\hat{o}_{i}$.} (a) Coefficients of each term in the effective Hamiltonian of a fluxonium qutrit with $E_J/E_C=2.78$ and $E_J/E_L=21.74$~\cite{vool2018driving} as a function of the external flux ($\Phi_\text{ext}$). Panel (b) is a zoom near the origin.  Coefficients of each term in the effective flux sine $\sin{(2\pi\phi/\Phi_0)}$ (c) and charge $q$ (d) operators for the fluxonium with $E_J/E_L=21.74$ as a function of $E_J/E_C$ at $\phi_{\rm ext}=0$. The legend in this graphics indicates which spin-1 operator accompanies each coefficient in the corresponding effective Hamiltonian or operator.}
\label{Fig:effectivehamiltonian}
\end{figure}

To keep computations tractable, we can orthogonalize the $\ket{L}$ and $\ket{R}$ states with respect to just the third excited state $\ket{-1} = \frac{\ket{L} - a_0 \ket{3}}{\sqrt{1-a^2}}$ and $\ket{1} = \frac{\ket{R} + a_0 \ket{3}}{\sqrt{1-a_0^2}},$ parameterizing the overlap with a new parameter  $a_0 = \braket{3|L}=-\braket{3|R},$ see Fig.\ \ref{Fig:eigenstates}(c). We will verify that this order of perturbation is sufficient to recover the qualitative form of the qutrit operators. Let us denote by $\pm\phi_\star$ the position of the local minima in units of flux, and assume that the capacitance of the fluxonium is large enough to prevent direct tunneling between them $\braket{L|R}=\braket{L|0}=\braket{R|0}\approx 0$. The projection of the  flux operator and of the Hamiltonian onto the qutrit spin-1 base $\left\{ \ket{-1}, \ket{0}, \ket{1}\right\}$ then reads:
\begin{align}
  \phi & { \approx} \widetilde{\phi}_\star S_z + b \left(S_xS_z +S_zS_x\right),\label{eq:phi_qutrit}\\
  H{_{\rm eff}} &{\approx} \epsilon S_z^2 +  \frac{\Delta}{2} \left(S_+^2+S_-^2\right)
  + \Phi_\text{ext} I_{0} \sin\left({\frac{2\pi\phi}{\Phi_0}}\right). \label{eq:op1}
\end{align}
  Here, $E_3$ is the eigenenergy of the third excited state, $\widetilde{\phi}_\star  = \frac{1-2 a_0^2}{1-a_0^2} \phi_\star,$ $b  = \frac{\sqrt{2}a_0}{\sqrt{1-a_0^2}}\braket{0|\phi|3}$ and $\Delta =  \frac{E_3a_0^2}{1-a_0^2}$ with $\epsilon=\mathcal{U}(\phi_\star)$, the potential energy of the system at the local minima at $\phi_{\rm ext}=0$.  Last but not least, we can compute the fluxonium charge operator acting on the qutrit subspace as in Ref.~\cite{hita2021ultrastrong}, via the Heisenberg equation $q = \frac{i C_f}{\hbar}[H,\phi]$:
\begin{align}
q {\approx} \frac{iC_f}{\hbar} \left[ ib\left(\epsilon-\Delta\right)S_y+
\Delta \widetilde{\phi}_\star \left(S_+^2-S_-^2\right)    \right].\label{eq:charge}
\end{align}

Our derivation expresses all operators in terms of the overlap $a$ and the energy $E_3$, quantities that may be estimated using the harmonic states of the right, left, and central wells. Intuitively, all terms containing $a$ in Eqs.~\eqref{eq:op1}-\eqref{eq:charge} are mediated by the third excited state. In the limit of large charging capacitance or \textit{heavy fluxonium}, the $\ket{-1},\ket{1}, \ket{0}$ states become strongly localized in left, center, and right well, making the factor $a$ exponentially small. In this case, the charge and flux operators converge respectively to $S_y$ and $S_z$ (notice the large factor $\epsilon$ in front of $S_y$ in the $q$ expansion), and the Hamiltonian is diagonalized by the states $\ket{0}$ and $\ket{\pm}=[\ket{L}\pm\ket{R}]/\sqrt{2}$ at zero bias $\Phi_\text{ext}=0$.  As explained in the introduction, in this limit the charge operator $q\approx S_y$ cannot mediate the decay of the $\ket{+}$ state to any of the other states that form the qutrit basis, neither the ground $\ket{0}$ or the $\ket{-}$ state, and the second excited state can be used to construct a BIC. It is important to note here that this is not true outside of this limit as, while direct transitions between the $\ket{+}$ and the ground $\ket{0}$ states are forbidden by symmetry, transitions to the $\ket{-}$ state can be mediated by the charge operator in Eq.\ \ref{eq:charge}. In this regime, this operator contains a non-negligible $(S^2_+-S_-^2)$ term which can mediate transitions from the $\ket{+}$ state to the $\ket{-}$ state, and eventually to the ground state $\ket{0}$ through the remaining    term in the charge operator $S_y$.

{ There is not a priori reason to expect the approximate expansion in Eqs.\ (\ref{eq:phi_qutrit}-\ref{eq:charge}) to capture the correct form of the relevant qutrit operators. Indeed, we have only imposed the orthogonalization of the qutrit subspace respect the 4th level, having thus an overlap with higher eigenstates. Nevertheless, we have obtained a good qualitative agreement between the \emph{exact} numerical diagonalizations and the operator expansions in those equations.} In our numerical approach, we compute the lowest energy eigenstates, project the relevant operators onto the qutrit basis and express them as a combination of 1-spin operators. For greater accuracy, we use the $\sin({\color{blue}2\pi\phi/\Phi_0})$ operator instead of $\phi$, and receive this and the $q$ operator as derivatives of the fluxonium's Hamiltonian with respect to flux and voltage perturbations, respectively. We use some of the subroutines of CirquitQ library\ \cite{aumann2021circuitq} and our code.

In Fig.~\ref{Fig:effectivehamiltonian} illustrates the excellent agreement between our predictions Eqs.~\eqref{eq:op1}-\eqref{eq:charge}, and the expansions of the Hamiltonian as a function of external magnetic flux $\Phi_\text{ext}=0$ (a-b), and of the flux (c) and charge (d) as a function of $E_J/E_C$ at $\Phi_\text{ext}=0$. Following our previous discussion, we see that charge and flux effectively become $S_x,S_y$ with other terms exponentially vanishing with increasing fluxonium capacitance. Knowing that at zero flux, the second excited state is $\ket{+}$, we have obtained rigorously that the matrix element of this state with the other qutrit state is suppressed exponentially fast for the heavy fluxonium. As we show next, this state becomes a BIC when the fluxonium is coupled via its charge operator to an extended object with a continuous spectrum.

\section{Bic in a fluxonium qutrit coupled to a waveguide}

Let us now discuss the dynamics of a fluxonium with a capacitive coupling $C_c$ to the continuum of propagating modes in a coplanar microwave guide, as shown in Fig.~\ref{Fig:setup}(a). From quantum optical considerations, the waveguide is a gapless medium supporting frequencies that would allow the fluxonium qubit to relax and decay from the $\ket{+}$ to the $\ket{0}$ or $\ket{-}$ states. However, based on our study of the charge operator, we conclude that the decay rate $\Gamma_{+0}=0$ due to flux reversal symmetry and that $\Gamma_{+-}$ becomes vanishingly small with increasing fluxonium capacitance. In this limit, at zero bias, the $\ket{+}$ state becomes a quasi-BIC state with an exponentially long lifetime\ \footnote{We use the term \textit{quasi-BIC} to indicate that the BIC state always has a finite lifetime.}

To compute the BIC's lifetime, we use the spin-boson model \cite{pino2018} for a  fluxonium connected to a waveguide of length $L$  and periodic boundary conditions. If the coupling is weak enough, the Hamiltonian of the combined system can be approximated as~\cite{garcia2015,parra2018quantum}:
\begin{align}
    H =& \frac{1}{2C_\Sigma}q^2+V(\phi)+\sum_{n=0}^{N-1} \hbar \omega_n \left(b_n^\dagger b_n+\frac{1}{2}\right) +\Delta H  \nonumber\\
    \Delta H =& \frac{C_c}{C_\Sigma}  q \sum_{n=0}^{N-1} (-1)^n\sqrt{\frac{\hbar\omega_n}{2c_0L}} i(b_n -b^\dagger_n). \label{eq:fullH}
\end{align}
We have expanded the waveguide Hamiltonian using the normal modes $[b_n, b_{m}]=i\hbar \delta_{n,m}$ with $n,m=0,\dots N-1,$ introducing waveguide's capacitance per unit length $c_0=C/L$ and the renormalized fluxonium's capacitance $C_\Sigma = C_f+C_c$.

\begin{figure}[t!]
\includegraphics[width=1\columnwidth]{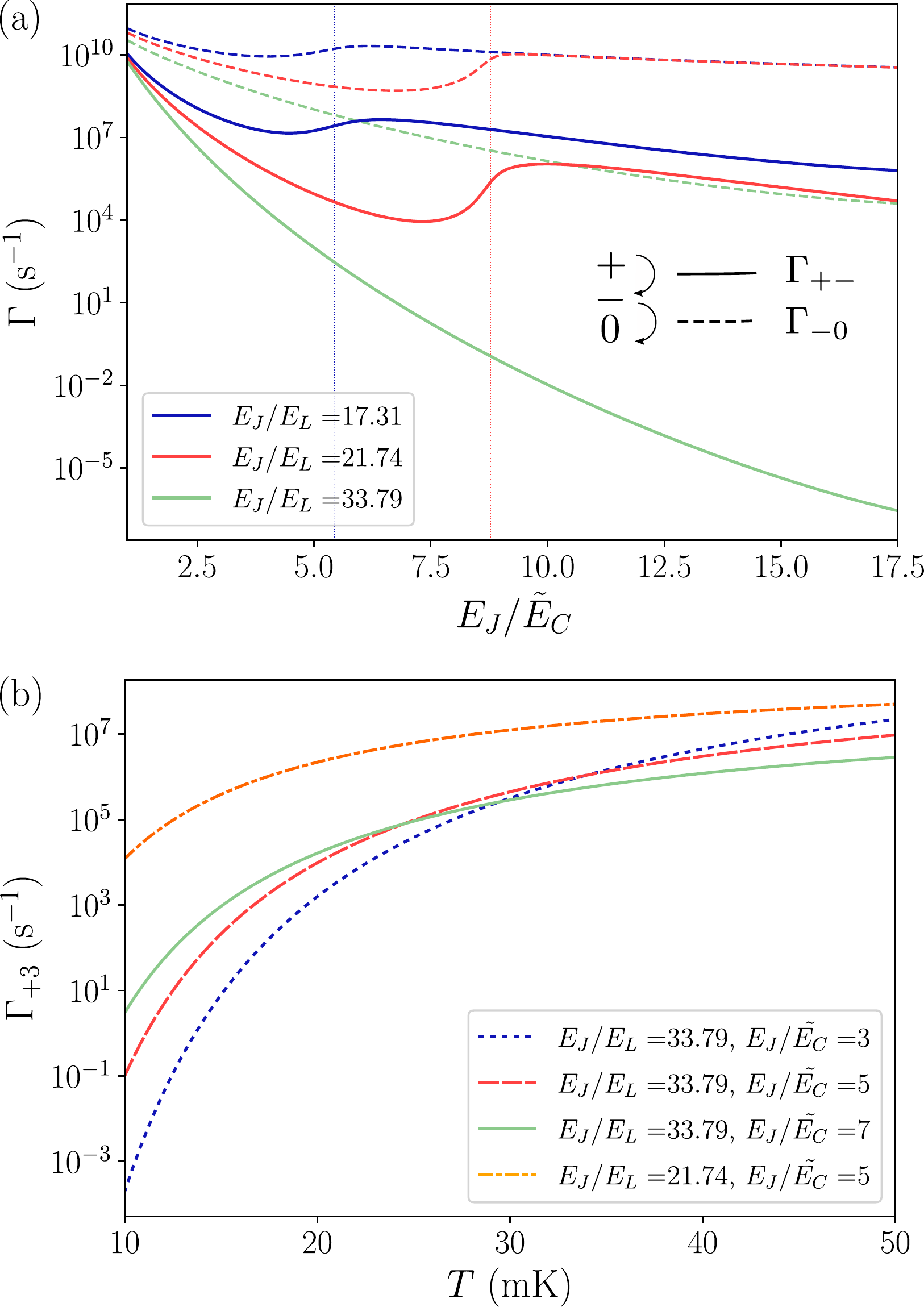}
\caption{(a) Transition rates for the allowed transitions in the fluxonium qutrit as a function of the ratio between Josephson and renormalized charging energies, $\widetilde{E}_C=\frac{e^2}{2C_{\Sigma}}$. The transition rates from the BIC state $\ket{+}$ to $\ket{-}$ appears as solid lines and from the $\ket{-}$ to $\ket{0}$ as dashed lines. The Josephson energy of the fluxonium is $E_J=10{\rm GHz}$ and the linear inductances are chosen so that they are experimentally realizable $E_J/E_L = 17.31$~\cite{manucharyan2009fluxonium}, $E_J/E_L = 21.74$~\cite{vool2014non} and $E_J/E_L = 33.79$~\cite{earnest2018}. The waveguide  has impedance $\mathcal{Z}=50\Omega$ and the coupling capacitance is taken $(E_C)_c=0.25{\rm GHz}.$ The vertical dashed lines indicate the value of $\widetilde{E}_C$ at which there is an avoided level crossing in the fluxonium spectrum between the second and third excited states. (b) Transition rates for the upwards transition between the $\ket{+}$ and $\ket{3}$ states of the fluxonium as a function of the temperature $T$.}
\label{Fig:BIClifetime}
\end{figure}

Fermi's Golden Rule~\cite{griffiths1962} is a good estimate for the transition rates $\Gamma_{ij}$ between fluxonium states $i$ and $j$ assisted by the modes of the waveguide. The formula requires the interaction Hamiltonian $\Delta{H}$ and the density of states $\rho(\omega)$ which, for a waveguide with a linear dispersion relation $\omega= \nu k$, is uniform $\rho(\omega)=L/(2\pi \nu)$. In this case Fermi's Golden Rule (see ~\cite{moskalenko2021} for an inductive coupling) predicts:
\begin{equation}
    \Gamma_{ij}=2\pi \left(\frac{C_c}{C_{\Sigma}}\right)^2 G_0\mathcal{Z} |\braket{i|N_f|j}|^2 \omega_{ij}, 
    \label{eq:fermi}
\end{equation}
as a function of the transition frequency $\omega_{ij}$, the number of Cooper pairs in the fluxonium $N_f=q/2e$ and the line's impedance $\mathcal{Z}=c_0\nu$. As discussed, direct transitions from $\ket{+}\rightarrow \ket{0}$ are forbidden by symmetry $\Gamma_{+0}=0$, and the $\ket{+}$ state can decay only via the $\ket{-}$ state. In other words, if we regard the Fermi's golden rule as a second order expansion in $\Delta H$ of the imaginary part of the self-energy\ \cite{stedman1971}, the main  contribution to that quantity at zero temperature are diagrams involving $\braket{+|\Delta H|\ -}$ matrix elements when $\Phi_{\rm ext} = 0$.

Fig.~\ref{Fig:BIClifetime}(a) displays the transition rates from the $\ket{+}$ to the $\ket{-}$ states (solid lines), as computed numerically using Fermi's Golden Rule and the exact eigenstates of the model. The parameters for each data set correspond to each of the experiments in Refs.\ \cite{manucharyan2009fluxonium,vool2014non,earnest2018}. The ideal lifetime of the BIC---the inverse of this relevant transition rate $T_\text{BIC}\approx 1/\Gamma_{+-}$---is exponentially enhanced as we increase the renormalized charging energy of the fluxonium $\widetilde{E}_C$. The other relevant transition rate in the qutrit subspace $\Gamma_{-0}$ (dashed line) decreases  much slower than $\Gamma_{+-},$ meaning that only the $+$ states is uncoupled from the waveguide. 

It is important to remark that for certain parameters, the energy of the third excited state $\ket{3}$ approaches the qutrit subspace and breaks all our approximations. As shown by the red curve for $E_J/E_L=21.74$ in Fig.~\ref{Fig:BIClifetime}(a), this manifests as a change in tendency in the decay rates, at a point where the BIC state ceases to exist. The value of $E_J/\widetilde{E}_C$ at which this happens depends strongly on the ratio $E_J/E_L$. This value determines the gap between the qutrit subspace and the high energy levels and sets a limit to the experimentally achievable lifetime of the BIC $\ket{+}$ state. However, as also shown in Fig.~\ref{Fig:BIClifetime}, appropriately tuning the ratio of inductive energies, such as $E_J/E_L=33.79$ with $E_J=10\,$GHz, yields lifetimes that are exponentially large, above $T_{1}\gg {\rm s}$, justifying that we call the $\ket{+}$ state a BIC.

\section{BIC's decay due to experimental imperfections}

We have already shown that an isolated quantum system composed of a heavy fluxonium and a waveguide displays the main properties of a BIC. However, we need to analyze other decay channels associated with the noise, which are always present in experiments. We consider two types of inelastic processes involving external degrees of freedom. In the first one, the fluxonium interchanges energy with the waveguide at a finite temperature, while, in the second, the energy is transferred to other environmental degrees of freedom. Considering all those processes, we will show that the BIC decay time is still rather long. We remark that our aim here is not to have a full description of the non-unitary dynamic of the system but to obtain a first estimation of the BIC decay time under realistic circumstances, together with a better understating of the main decay channels. When computing numerical results in this section, we will employ the fluxonium parameters used in each of the experiments in Refs.\ \cite{manucharyan2009fluxonium,vool2014non,earnest2018}, while noise parameters are obtained from the ones reported in Refs.\ \cite{manucharyan2009fluxonium, earnest2018, Zhang2021}.

\subsection{Temperature and noise assisted BIC's decay into the waveguide}\label{sec:decaywg}

We start with the effect of a non-zero temperature in the waveguide, which could assist transitions from the BIC to higher energy states. We only consider the transitions from the BIC state $\ket{+}$ to the third excited state $\ket{3},$ as higher energy states are more difficult to be excited. We show in Fig.~\ref{Fig:BIClifetime}(b) the transition rates between $\ket{+}$ and $\ket{3}$ excited states as a function of temperature $T,$ computed with a version of Eq.~\ref{eq:fermi} which takes into account the finite probability of thermal photons in the waveguide. We notice that the BIC lifetime is strongly reduced at temperatures larger than the gap between the BIC and third excited state. It is because the relevant matrix elements between the $\ket{+}$ and $\ket{3}$ states always overlap significantly, even in the heavy fluxonium limit. Thus we should choose fluxonium parameters for which those two states have an energy gap larger than the temperature. This is indeed the case for the curves in Fig.~\ref{Fig:BIClifetime}(b) corresponding to $E_J/E_L\approx33$, which  displays transition rates with lifetimes of the order of $T_1\approx 1$ms at an experimentally realizable temperature of $T<15$mK.

The next  decay channel we analyze is spontaneous photon emission into the waveguide at a non-zero flux. Rigorously, the symmetry that protects the BIC state only appears for integer flux values $\Phi_\text{ext} = \Phi_0\times\mathbb{Z}$. However, flux noise---specially the slow one---can create a flux bias for long enough time so that the BIC decays to the ground. To understand this, we rely on state-of-the-art models for how low-frequency flux noise penetrates Josephson devices in quantum information applications~\cite{koch1983flicker,Bialczak2007, Paladino2014}. This noise has a power spectrum that can be approximated as 
\begin{equation}
 S^\Phi(\omega)\approx 2\pi A^2/\omega\label{eq:1fnoise}
\end{equation}
with $A\approx (10^{-5}-10^{-6})\Phi_0$~\cite{koch2007model, yan2016, kumar2016, nguyen2019}, which implies quasi-static fluctuations with an amplitude $\sigma\sim (10^{-5}-10^{-6})\Phi_0$ (see Supplemental Material). Fig.~\ref{Fig:breakingtheBIC}(a) displays the transition rate from the BIC to the ground state as a function of the flux deviation $\Phi_\text{ext}\neq 0$. This figure also confirms that the BIC becomes more sensitive to external flux perturbations as the ratio $E_J/E_C$ is increased, so we should not make a "too heavy fluxonium". The expected low-frequency flux fluctuations, with amplitude $10^{-5}-10^{-6}\Phi_0$, are denoted by a colored region in Fig.~\ref{Fig:breakingtheBIC}(a). An example of parameters that work well corresponds to the purple solid line from Fig.~\ref{Fig:breakingtheBIC}(a) with $\frac{E_J}{E_C}\approx 5,\ \frac{E_J}{E_L}\approx 30 $. For a pessimistic estimation of the fluctuations $10^{-5}\Phi_0,$ we would obtain a decay time $T_{1}\sim 10^{-1}\,\text{ms}$, while for a moderately optimistic noise amplitude $ 10^{-6}\Phi_0,$ the expected decay time is $T_1\sim 10{\rm ms},$ of the same order of the radiative losses from $\ket{+}\rightarrow \ket{-}$ in Fig.\ \ref{Fig:BIClifetime}.

\begin{figure}[t!]
\includegraphics[width=1.\columnwidth]{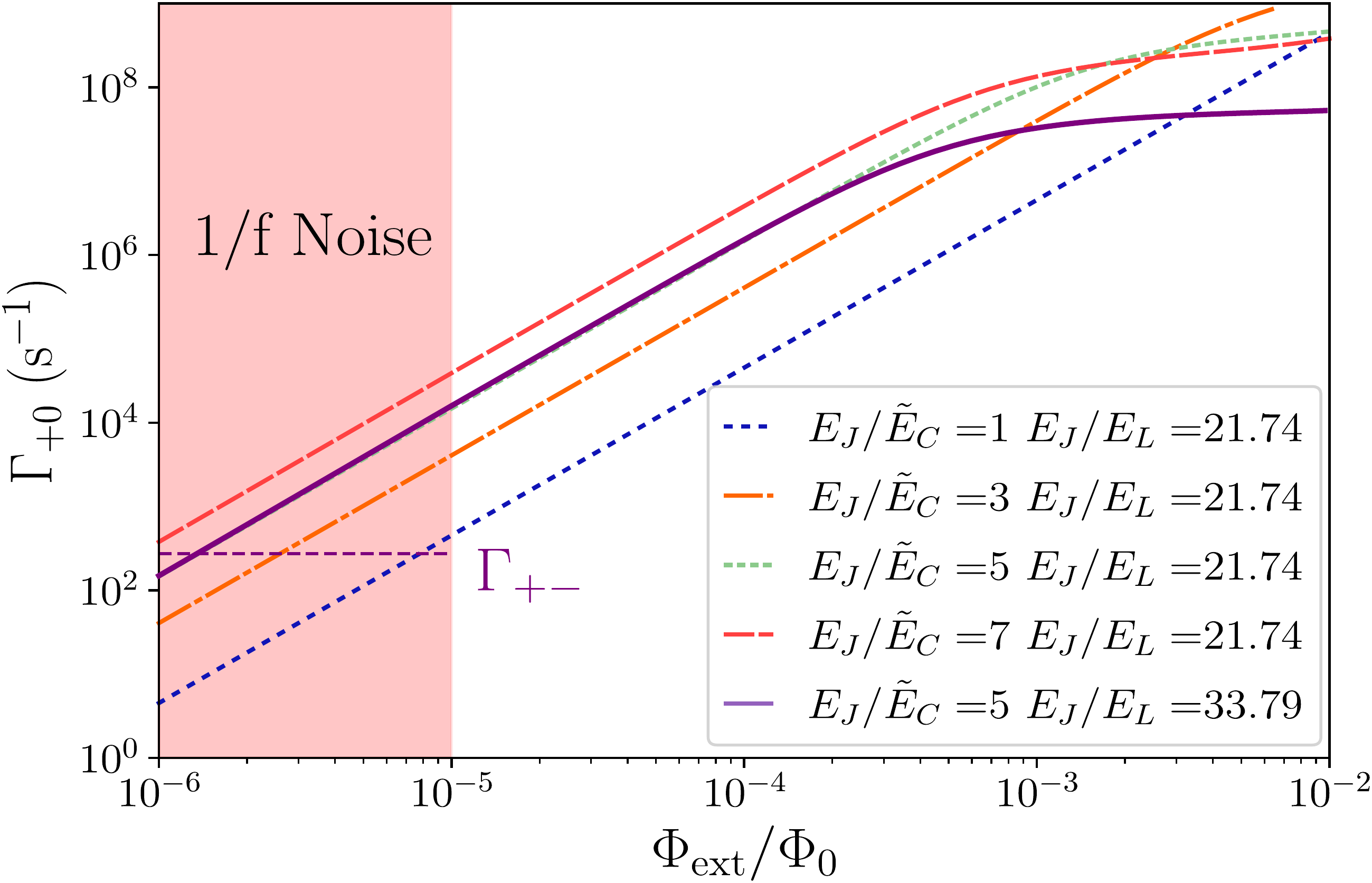}
\caption{ Transition rate from $\ket{+}$ to $\ket{0}$ due to radiative losses to the waveguide as a function of the external flux, $\Phi_\text{ext}$, in log-log scale. The Josephson energy of the fluxonium is $E_J=10{\rm GHz}$. The waveguide  has impedance $\mathcal{Z}=50\Omega$ and the coupling capacitance is taken $(E_C)_c=0.25{\rm GHz}.$ The horizontal dashed line signals the transition rate $\Gamma_{+-}\approx 3 *10^2 {\rm Hz}$ for the fluxonium parameters of the purple solid line $E_J/\widetilde{E}_C=5$ and $E_J/E_L=33.79.$ (see Fig.~\ref{Fig:BIClifetime}). The coloured region signals the typical amplitudes of quasi-static fluctuations in fluxoniums produced by $1/f$-noise.}
\label{Fig:breakingtheBIC}
\end{figure}

\subsection{BIC's decay into the environment}

Besides fluxonium relaxation due to the coupling with the waveguide, the BIC may also decay by releasing  or absorbing energy to the fluxonium environment. Due to our simplified circuit design, the noise seen by the fluxonium is dominated by slow $1/f$-flux noise and dielectric and inductive losses~\cite{Zhang2021,nguyen2019}. We quantify their effect in the BIC decay time using again Fermi's golden rule\ \cite{Zhang2021,Clerk2010}. For a noise source with amplitude $f(t)$ that is coupled to the fluxonium via operator $\hat{O},$ giving and interaction of the form $\Delta H=f(t)\hat{O}$, the transition rate from the $i$ to $j$ states is:
\begin{equation}
\Gamma_{ij}^{\rm noise}=\frac{1}{\hbar^2}\left|\bra{j}\hat{O}\ket{i}\right|^2S(\omega_{ij})
\label{eq:noise}
\end{equation}
where $S(\omega)=\int^{\infty}_{-\infty}\left<f(t)f(0)\right>e^{i\omega t}dt$ is the noise source's spectral density that can be determined following the fluctuation-dissipation theorem ~\cite{coleman2015}, and $\omega_{ij}$ is the transition frequency.  We use the positive  frequency component of the spectral density $S(\omega)$  to study BIC decay and the negative one $S(-\omega)$ for BIC transition upwards, where both components are related by $S(-\omega)=S(\omega)e^{-\frac{\hbar\omega}{K_BT}}$. We remind that direct transitions from the BIC to the ground are forbidden if the operator that couples to the noise obeys flux reversal symmetry, as it is the case in what follows.

\begin{figure}[t!]
\includegraphics[width=1.\columnwidth]{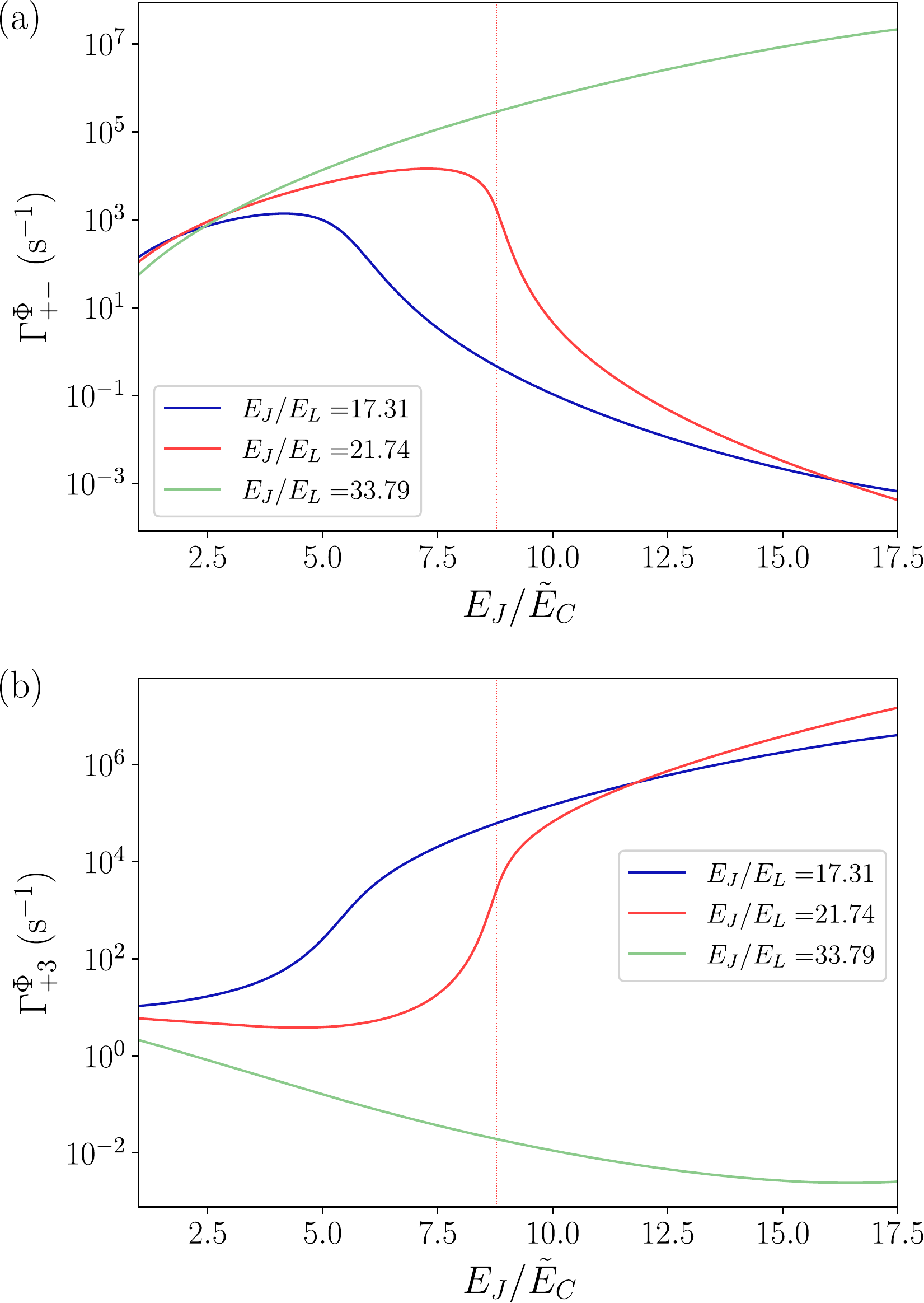}
\caption{Transition rates for the $1/f$-flux noise induced transitions in the fluxonium as a function of the ratio between Josephson and renormalized charging energies, $\tilde{E}_C =\frac{e^2}{C_{\Sigma}}$. The Josephson energy of the fluxonium is $E_J = 10$GHz and the linear inductances are chosen so that they are experimentally realizable $E_J/E_L = 17.31$~\cite{manucharyan2009fluxonium}, $E_J/E_L = 21.74$~\cite{vool2014non} and $E_J/E_L = 33.79$~\cite{earnest2018}. The $1/f$-flux noise amplitude is taken to be $A=5\times10^{-6}\Phi_0$. The vertical dashed lines in both plots indicate the value of $\widetilde{E}_C$ at which there is an avoided level crossing in the fluxonium spectrum between the second and third excited states.  (a) Transition rates for the $1/f$-flux noise induced transition from the BIC state $\ket{+}$ to $\ket{-}$. (b) Transition rates for the $1/f$-flux noise induced transition from the BIC state $\ket{+}$ to $\ket{3}$. }
\label{Fig:flux noise}
\end{figure}

Apart from inducing decay to the waveguide, $1/f$-flux noise couples to the fluxonium persistent current ${I}_p={\phi}/L$ and produce its relaxation/excitation at a transition rate given by Eq.~\ref{eq:noise} with spectral density given in Eq.\ \ref{eq:1fnoise}. Fig.~\ref{Fig:flux noise}(a) shows the numerically computed $\ket{+}\rightarrow\ket{-}$ transition rates for an intermediate 1/f-noise amplitude $A=5\times 10^{-6}\Phi_0$. Contrary to what happens when studying the decay transitions assisted by the waveguide, the $\ket{+}\rightarrow\ket{-}$ transition rate grows as the charging energy of the fluxonium increases. For large values of this parameter, the $1/f$-flux noise dominates the transition time to the $\ket{-},$ significantly reducing the lifetime of the BIC state, so that the regime of a very heavy fluxonium should be avoided due to its enhanced sensibility to the $1/f$-noise. This enhancement occurs because the exponential decrease of the gap $E_{+-}$ with $E_J/E_C$ produces an exponential increase of the effective noise that affects the decay rate. We have also analyzed upward transitions at $T\neq 0$ due to $1/f$-noise in  
Fig. ~\ref{Fig:flux noise}(b), which yields smaller transition rates than the previous one $\ket{+}\rightarrow \ket{-}.$

\begin{figure}[t!]
\includegraphics[width=1.\columnwidth]{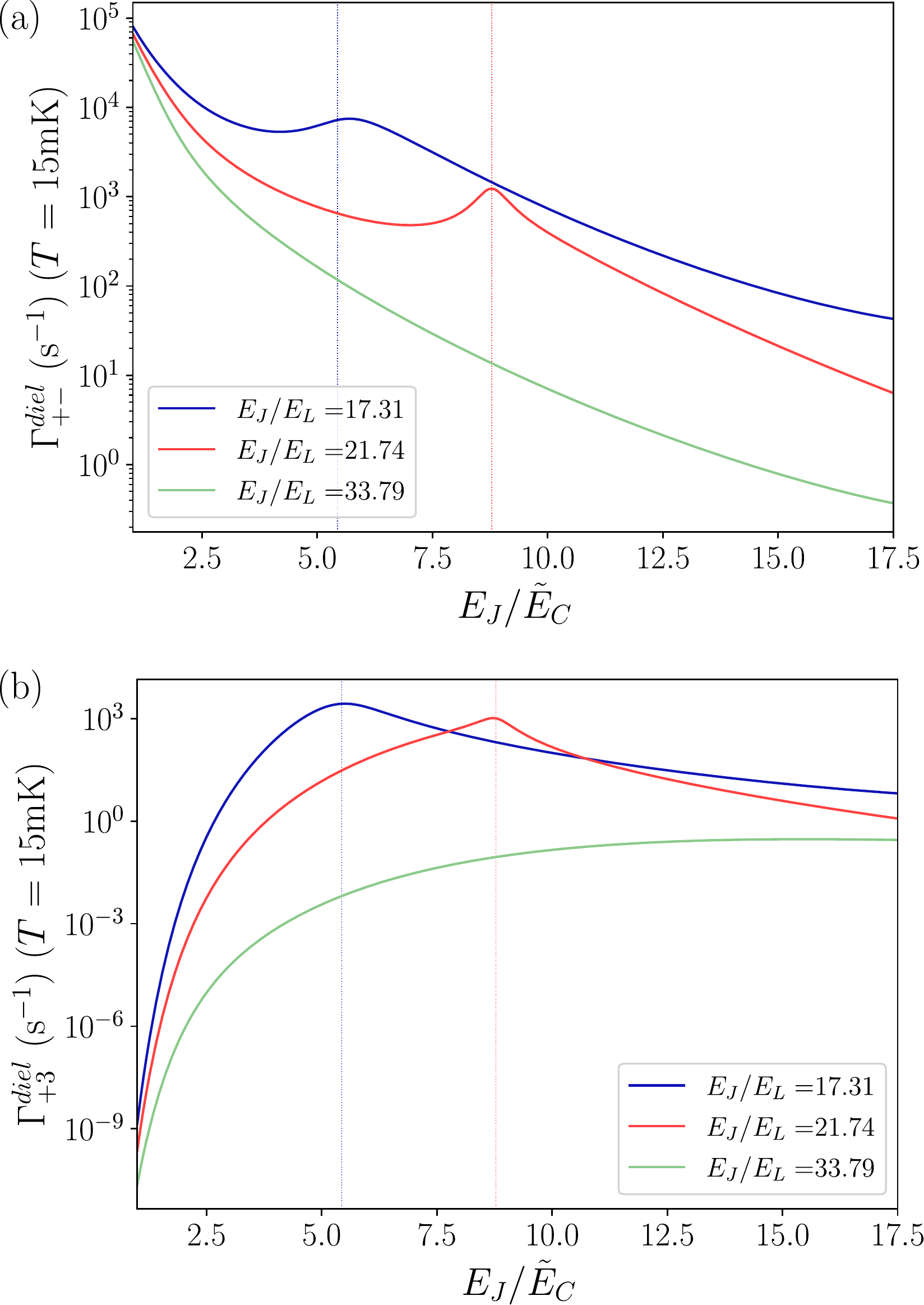}
\caption{Transition rates for the dielectic noise induced transitions in the
fluxonium at $15$mK as a function of the ratio between Josephson
and renormalized charging energies, $\tilde{E}_C =\frac{e^2}{C_{\Sigma}}$. The Josephson energy of the fluxonium is $E_J = 10$GHz and the linear inductances are chosen so that they are experimentally realizable $E_J/E_L = 17.31$~\cite{manucharyan2009fluxonium}, $E_J/E_L = 21.74$~\cite{vool2014non} and $E_J/E_L = 33.79$~\cite{earnest2018}. The dielectric quality factor is approximated to be $Q_{\rm diel}=1/(4\times10^{-6})$ ~\cite{Zhang2021}. The vertical dashed lines in both plots indicate the value of $\widetilde{E}_C$ at which there is an avoided level crossing in the fluxonium spectrum between the second and third excited states.  (a) Transition rates for the dielectric noise induced transition from the BIC state $\ket{+}$ to $\ket{-}$. (b) Transition rates for the dielectric noise induced transition from the BIC state $\ket{+}$ to $\ket{3}$. }
\label{Fig:dielectricnoise}
\end{figure}

It is argued in Ref.~\cite{nguyen2019} that one of the main sources of relaxation in the fluxonium qubit is tangential losses into the dielectric, similarly as in phase  or transmons qubits~\cite{martinis2005, martinis2022}. Dielectric losses  can be identified as current noise, coming from the resistive part of the shunting capacitor, that couples to fluxonium flux $\phi$~\cite{Zhang2021}. Its spectral density is $S^{\rm diel}(\omega)=\frac{\hbar\omega^2C}{Q_{\rm diel}}\left[1+\coth{\left(\frac{\hbar\omega}{2K_BT}\right)}\right]$, being $1/Q_{\rm diel}$ the loss tangent of the shunting capacitor which is proportional to its impedance.  Following Eq.~\ref{eq:noise}, we compute the relaxation rates for the dielectric noise induced transitions $\ket{+}\rightarrow \ket{-}$  and also $\ket{+}\rightarrow \ket{3}$ in panels (a) and (b) of Fig.~\ref{Fig:dielectricnoise}, respectively. We assume a fixed value of $Q_{\rm diel}= 1/(4\times 10^{-6})$ as in Ref.~\cite{Zhang2021} and a temperature of $15$mK.  Although the dielectric losses may dominate over the natural decay in the waveguide, it is still the $1/f$-flux decay from previous paragraph the one that seems more problematic, at least in the parameter regime we are working. The $\ket{+}\rightarrow\ket{3}$ transition rates are at most of the same order of magnitude of the $\ket{+}\rightarrow\ket{-}$ transition rates, and, hence, do not notably modify the lifetime of the BIC state.

\begin{figure}[t!]
\includegraphics[width=1.\columnwidth]{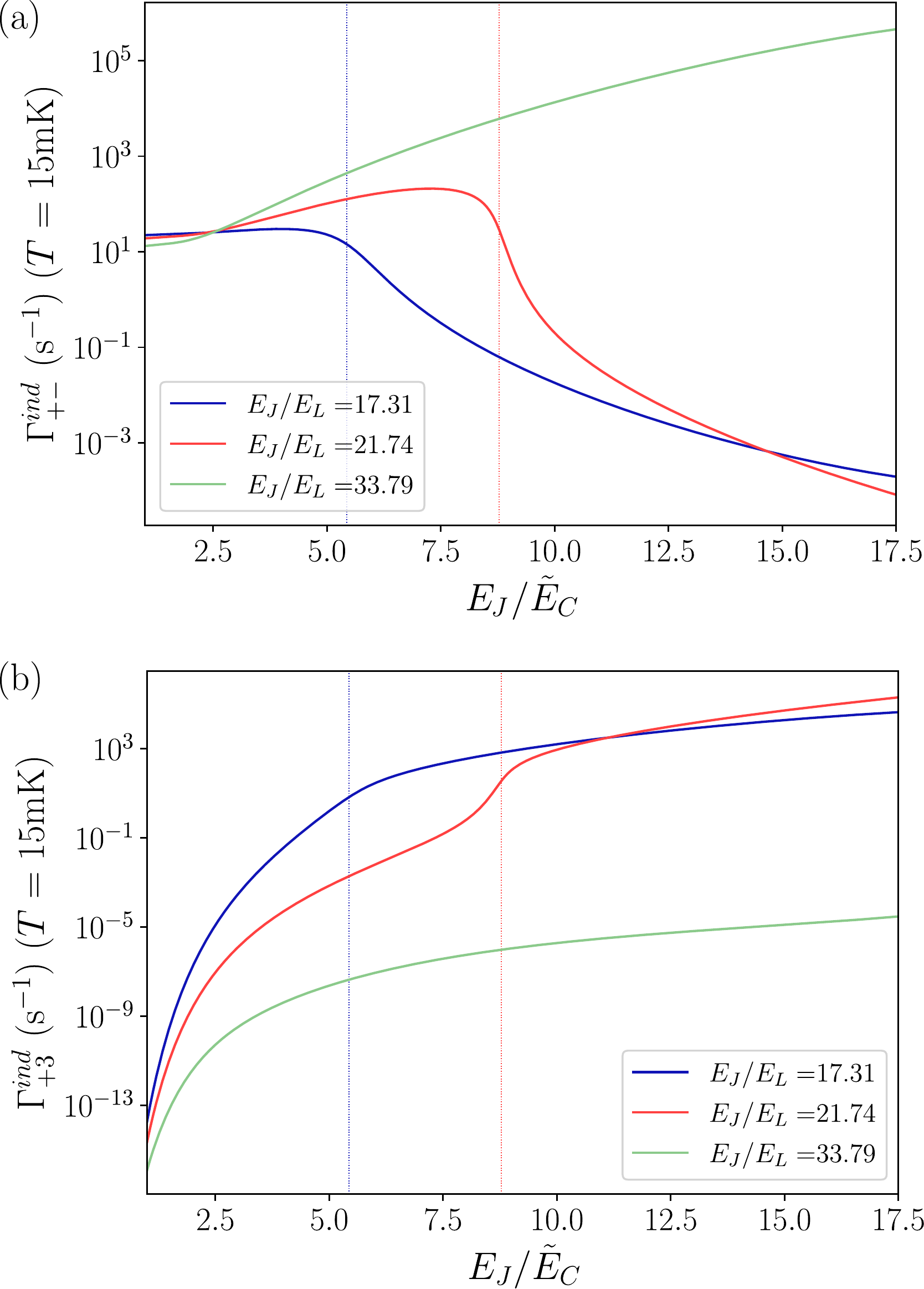}
\caption{Transition rates for the inductive noise induced transitions in the
fluxonium at $15$mK as a function of the ratio between Josephson
and renormalized charging energies, $\tilde{E}_C =\frac{e^2}{C_{\Sigma}}$. The Josephson energy of the fluxonium is $E_J = 10$GHz and the linear inductances are chosen so that they are experimentally realizable $E_J/E_L = 17.31$~\cite{manucharyan2009fluxonium}, $E_J/E_L = 21.74$~\cite{vool2014non} and $E_J/E_L = 33.79$~\cite{earnest2018}. The inductive quality factor is approximated to be $Q_{ind}=8\times10^9$ ~\cite{Zhang2021}. The vertical dashed lines in both plots indicate the value of $\widetilde{E}_C$ at which there is an avoided level crossing in the fluxonium spectrum between the second and third excited states.  (a) Transition rates for the inductive noise induced transition from the BIC state $\ket{+}$ to $\ket{-}$. (b) Transition rates for the inductive noise induced transition from the BIC state $\ket{+}$ to $\ket{3}$. }
\label{Fig:inductivenoise}
\end{figure}

Another important relaxation mechanism that affects the fluxonium is inductive losses. It can be thought as a current noise, but this time associated with the resistive part of the inductor, that couples to the flux operator $\phi$ with spectral density $S^{\rm ind}(\omega)=\frac{\hbar}{LQ_{ind}}\left[1+\coth{\left(\frac{\hbar\omega}{2K_BT}\right)}\right]$ ~\cite{Zhang2021}. 
In Fig. ~\ref{Fig:inductivenoise}, we show the relaxation rates associated with this mechanism using a quality factor $Q_{\rm ind}=8\times 10^{9}$ ~\cite{Zhang2021} at $15$mK . Similarly to what happens for the $1/f$-noise, the relaxation ratios increase when the fluxonium's charging energy increases, although their actual values are smaller than the ones we saw for the case of $1/f$-noise, making the latter dominate the decay of the BIC. Transitions between the second $\ket{+}$ and the third $\ket{3}$ excited states are irrelevant for this type of noise since they yield significantly low transition rates.

We have summarized the decay time of all the previous mechanisms in table\ \ref{tab:table}. First,  upwards transitions in the waveguide are rather dangerous due to a finite temperature. It is because the matrix elements between the BIC and the third excited state are not small, not even within the limit of a very heavy fluxonium. Thus, we will need to set a temperature smaller than the distance between those two levels in order to avoid these processes. We have found that a temperature of  $15$mK is low enough for our parameters so that other mechanisms dominate the BIC decay process. Between them, the ones causing the fastest decay are related to $1/f$ flux noise. Indeed, we have learned that the task of finding optimal parameters to increase BIC's lifetimes in the open system is not as easy as for the closed one. The trick of increasing the mass of the fluxonium, which could yield astronomically large BIC's decay times when fully isolated, does not work in the open system as it enhances considerably the effect of $1/f-$flux noise.

\begin{table*}
\caption{\label{tab:table}Transition rates corresponding to all the mechanisms that determine the lifetime of the quasi-BIC at $15$mK. The three first columns correspond to fluxonium parameters, Josephson energy $E_J=10{\rm GHz}$ and ratios of this quantity with renormalized charging energy and linear inductances. The ratios $E_J/E_L = 21.74$ and $E_J/E_L = 33.79 $ are the ones in Refs.~\cite{vool2014non} and~\cite{earnest2018}, while the renormalized charging energy $E_J/\tilde{E}_C$ is chosen to offer reasonable lifetimes. The decay rates without superscript corresponds to decay into the waveguide at a non-zero temperature (downwards and upwards). The ones with superscript $\Phi_{\rm ext}\neq 0$ are due to decay into the waveguide due to a finite flux bias produced by $1/f$-flux noise. The other decay rates are due to relaxation to the environment, the superscript $\Phi$ corresponding again to $1/f$-flux noise and the other to dielectric and inductive losses. We use a temperature of $15$mK, an intermediate value of the $1/f$-flux noise amplitude, $A=5\times10^{-6}\Phi_0$, and fixed quality factors,  $Q_{\rm diel}=1/(4\times10^{-6})$ and $Q_{ind}=8\times10^9,$ extracted from Ref.~\cite{Zhang2021}.  
}
\begin{ruledtabular}
\begin{tabular}{c c c c c c c c c c c c c c c}
    $E_J$ &     $\frac{E_J}{\tilde{E}_C}$ &     $\frac{E_J}{E_L}$ &     $\Gamma_{+-}$ &     $\Gamma_{+3}$ &     $\Gamma_{+0}^{\Phi_{\rm ext}\neq0}$ &     $\Gamma_{+-}^{\Phi_{\rm ext} \neq 0}$ &     $\Gamma_{+3}^{\Phi_{\rm ext} \neq 0}$ &     $\Gamma_{+-}^{\Phi}$ &     $\Gamma_{+3}^{\Phi}$  &     $\Gamma_{+-}^{\rm diel}$ &     $\Gamma_{+3}^{\rm diel}$ &     $\Gamma_{+-}^{\rm ind}$&     $\Gamma_{+3}^{\rm ind}$  &     $T_1$\\ 
    (GHz) &  & &      (s$^{-1}$) &     (s$^{-1}$)  &     (s$^{-1}$)  &     (s$^{-1}$)  &     (s$^{-1}$)  &     (s$^{-1}$) &     (s$^{-1}$)&     (s$^{-1}$) &     (s$^{-1}$) &     (s$^{-1}$)&     (s$^{-1}$)&     (ms) \\ 
\hline
       $10$   &       $5$                &          $21.74$             &           $1\times10^{5}$ &           $4\times10^{5}$  &           $4\times10^{3}$  &           $1\times10^{5}$  &           $90$ &           $7\times10^{3}$ &           $4$ &           $8\times10^{2}$&           $20$   &           $1\times10^{2}$ &           $7\times10^{-4}$   &           $2\times10^{-3}$           \\ 
       $10$   &       $5$                &          $33.79$             &           $1\times10^{3}$ &           $2\times10^{2}$  &           $1\times10^{3}$  &           $1\times10^{3}$  &           $2\times10^{2}$ &           $1\times10^{4}$ &           $0.2$ &           $2\times10^{2}$&          $3\times10^{-3}$   &           $3\times10^{2}$ &           $2\times10^{-8}$   &           $5\times10^{-2}$  \\
\end{tabular}
\end{ruledtabular}
\end{table*}

\section{State preparation}\label{sec:sp}

The simplest strategy to create the BIC state may be to suppress the BIC protection by a small flux bias $\delta \Phi_{\rm ext}\approx 10^{-3}\Phi_0$  and then populate it with an appropriate driving field. From the results in Fig.\ \ref{Fig:breakingtheBIC}, we see that a pulse of around $10{\rm ns}$ would be enough. Once this is done, we need to return to zero external flux to restore the BIC protection. We can give a rough estimation of the time $\Delta t$ required to do so adiabatically by applying the Landau-Zener formula to the subspace expanded by $\ket{+},\ket{-}$. The non-adiabatic transition are suppressed by a exponential factor $\exp{(-2\pi \Gamma_{\rm LZ})},$ where we approximate $\Gamma_{\rm LZ}\approx a^2\Delta t/(\hbar \Delta E).$ The  $a$ in the previous formula is the energy difference at the avoided level crossing, and $\Delta E$ is the energy difference between the $\ket{+}$ state at zero and at a small bias flux $\delta \Phi_{\rm ext}$. We have numerically estimated those quantities obtaining that an approximate time of $\Delta t\approx 10^2{\rm ns}$ is needed for adiabaticity. Related to this, Ref.\ \cite{Zhang2021} reports on a similar protocol, where diabatic Landau-Zener transitions are employed to operate the qubit subspace of a heavy fluxonium at $\Phi_{\rm ext}=0$ (which rather similar to our $\ket{+},\ket{-}$ subspace) at typical times of $10^2$ns\ \cite{Zhang2021}.

Another possibility is to use a non-linear coupling in fluxonium charge or phase operators to drive the otherwise forbidden transition $\ket{0}\rightarrow \ket{+}.$ Similar ideas have been experimentally demonstrated for a more difficult set-up where a non-linear coupler between a fluxonium and a resonator allowed to drive symmetry forbidden transition of the full system\ \cite{vool2018driving}. In our proposed experiment, a non-linear coupling in the fluxonium flux is already present when coupling to the waveguide through a mutual inductance (Eq.\ \ref{eq:fullH}  with $\Phi_{\rm ext}=M I_w$ being the flux created by the waveguide). However, the residual mutual inductance is probably too small, so the non-linear terms are negligible. Although it is beyond the scope of this work to analyze the inductive coupling,  it could provide an interesting path to create the BIC for strong couplings\ \cite{forn2017, yoshihara2017}. If needed, one could try to enhance those non-linearities and make the coupling tunable via a SQUID, similar to what we have proposed in Ref.\ \cite{hita2021three} for flux qubits. 

One could also create the BIC using Raman transitions, similar to the experimental work in Ref.\ \cite{earnest2018}. The idea is to use the third excited state, the first one out of the qutrit subspace, to produce a transition scheme $\ket{0}\rightarrow \ket{3}\rightarrow \ket{+}$ at zero external flux. This can be done by stimulating the transition $\ket{3}\rightarrow \ket{+}$ with a probe field while pumping at the frequency of the $\ket{0}\rightarrow \ket{3}$ transition. In Ref.\ \cite{earnest2018}, they showed that this method allowed to create of a state mainly localized at the right potential well of a rather heavy fluxonium $E_{\rm J}/E_{\rm C}\sim 18$ in a time around $400 {\rm ns}.$

\section{Conclusions}

In summary, we have shown how to construct a compact BIC living in a superconducting fluxonium qutrit capacitively connected to an open microwave guide. This device can be brought to a regime where the second excited state is a quasi-BIC one, displaying a long decay time that, in the ideal case, can reach up to seconds. The critical ingredient for this BIC is the destructive quantum interference between opposite persistent current states that appears at $\Phi_{\rm ext }=0$\ \cite{ahumada2018bound}. As a result, we have a fully tunable BIC that can be brought in and out of the protected state by tuning the magnetic field in its loop. 

We have carefully analyzed several noise mechanisms, obtaining that noise limits the BIC state's lifetime, especially for large values of the fluxonium's charging energy. However, choosing the right parameters of the system could enable obtaining fairly large BIC lifetimes $T_{1}\approx 10^{-1}$ms (see Table \ref{tab:table}), much larger than preparation times which we estimated in the range of $10^2$ns (Sec. \ref{sec:sp}). Regarding the different noise mechanisms, we have seen that elastic processes where BIC gets excited to a higher level are detrimental to its lifetime. However, they can be suppressed by increasing $E_J/E_L$ or decreasing temperature. Once this is done, $1/f$-noise is likely to dominate the relaxation of the BIC, contrary to what has been seen in previous fluxonium qubits where decay was mainly induced by dielectric losses\ \cite{nguyen2019,Zhang2021}.  Added to elastic processes with the $1/f$ degrees of freedom, this noise can produce a long flux bias removing the symmetry protection of the BIC. It may be the case that this new type of decay channel involving $1/f-$noise plays an essential role in other persistent current qubits operated at "sweet spots." In any case, it would be highly beneficial for the BIC to perform an appropriate surface treatment of the fluxonium to reduce $1/f$-noise as in Ref.\ \cite{kumar2016}.

The possibility of creating long-lived BIC states in small fluxonium devices is exciting as a scalable platform for storing protected quantum information. However, there is a compromise between the lifetime of the BIC state and the possibility of accessing those states using external fields, as discussed above. We have commented on several ways of  engineering the BIC $\ket{+}$ states by dynamically tuning the external magnetic fields while controlling the injection of photons, using multi-photon induced transitions via excited states or via a non-linear fluxonium-waveguide coupler, as explained in Sec.\ \ref{sec:sp}.

We have also shown that the fluxonium BIC configuration is sensitive to external magnetic fields. In particular, we believe that it is possible to build a  magnetic field sensor by monitoring the $\ket{+}\leftrightarrow\ket{0}$ resonance, as both the intensity and linewidth of that resonance depend on small deviations of the magnetic flux experienced by the fluxonium [cf. Fig.~\ref{Fig:breakingtheBIC}]. For instances, a change in magnetic field of $\Delta B= 10 {\rm nT}$ would produce a change in the magnetic flux of  a typical fluxonium (area $\sim 10^3 \mu{\rm m}$\ \cite{nguyen2019}) of $\Delta \Phi \approx 5\ 10^{-3}\Phi_0,$ which activates the BIC. The activation associated with the excess magnetic field can then be recorded in the resonance measurements. We believe this method could be competitive compared to others, as the ones based on nitrogen-vacancy centers\ \cite{kuwahata2020magnetometer}.

\par

\begin{acknowledgments}

The authors would like to thank the anonymous referees who provided useful and detailed comments on a previous version of the manuscript. We would also like to thank 
Ioan Pop for pointing us out the importance of upwards transitions in BIC's decay time. This work is supported by the European Commission FET-Open project AVaQus GA 899561, by Proyecto Sinérgico CAM 2020 Y2020/TCS-6545 (NanoQuCo-CM) and CSIC Quantum Technologies Platform PTI-001. We acknowledge Centro de Supercomputación de Galicia (CESGA) who provided access to the supercomputer FinisTerrae for performing simulations. Part of the simulations were performed on the Tirant node of the Red de Española de Supercomputación (RES). M. P. acknowledges support by Spanish MCIN/AEI/10.13039/501100011033 through Grant No. PID2020-114830GB-I0. P.A.O. acknowledges support from FONDECYT Grant No. 1180914 and 1201876.
\end{acknowledgments}

\appendix

\section{Quasi-Static noise}

Typical flux fluctuations for a fluxonium device can be extracted from its noise power spectrum. We are interested in the low-frequency flux noise, which can bias the device for long enough time and, thus, produce a decay of the BIC into the waveguide. The dominant source of low-frequency flux noise in the type of devices treated here is the one referred as $1/f$-noise, so it is important to understand its properties in order to characterize BIC decay in realistic situations. 

We assume a power spectrum as explained in the Appendix of Ref.~\cite{nguyen2019}, based on many previous experimental results: 
$$S(\omega) =2\pi A^2/\omega$$ 
with $A=(10^{-5}-10^{-6})\Phi_0.$ We set a low and high frequency cut-off  for the $1/f$-noise as $\gamma_-=10^{-2}{\rm Hz}$ and $\gamma_+=10^1{\rm Hz},$ which is consistent with the experimental results in Ref.~\cite{yan2016}  (red points in Fig. 3 of that reference).  We will see that, in any case, the values of those cut-offs do not affect that much the final results. 

Once the noise model is set, we can extract the fluctuations at the low frequency cut-off, the important one for the BIC, as the real part of the Fourier transform of the power spectrum:
\begin{align}
    \sigma^2 \approx \av{\Phi_\text{ext}(t)\Phi_\text{ext}(t+\tau)} =A^2 \int_{\gamma_-}^{\gamma_+} \frac{d\omega}{\omega} 
    \cos\left( \omega \tau\right).
\end{align}
The time $\tau\sim 1/\gamma_-$ should be of the order of the inverse of the low-frequency cut-off so to get the amplitude of the quasi-static fluctuations. Setting this value in the previous formula, we obtain:
\begin{align}
    \sigma^2 \approx A^2 \int_{1}^{\frac{\gamma_+}{\gamma_-}} \frac{dx}{x} 
    \cos\left(x\right).
\end{align}
Expressing the previous results in terms of the cosine integral function 
\begin{equation}
    {\rm Ci(x)}=- \int_{x}^{\infty} \frac{dx}{x}     \cos\left(x\right),
\end{equation}
and taking $\frac{\gamma_+}{\gamma_-}=10^{3}$ we get the desired amplitude as given by:
\begin{align}
    \sigma^2 = A^2 \left[ - {\rm Ci}(10^{3}) + {\rm Ci}(1)\right] \sim \mathcal{O}(1) A^2.
\end{align}
As we previously state, this result do not depend strongly on the low and high-energy cut-offs as $|{\rm Ci}(x)|< 1/x$ vanishes fast.  For our purposes, we take $\sigma \approx A,$ because the factor of order one is irrelevant due to the uncertainty in the magnitude $A$ itself. Thus, we have taken the fluctuations in Fig. (4) of the main text as given by $A=(10^{-5}-10^{-6})\Phi_0.$

\bibliography{annealing}% Produces the bibliography via BibTeX.

\end{document}